\documentclass[12pt, a4paper]{article}

\usepackage{amsmath,amssymb,amsfonts}
\usepackage{graphicx,esint}
\usepackage[colorlinks,hypertexnames=false]{hyperref}
\usepackage{cite}

\begin{document}

\title{Analytic calculation of dynamical friction for Plummer sphere in ultralight dark matter}
\author{O.V.~Barabash${}^1$, T.V.~Gorkavenko${}^1$,
V.M.~Gorkavenko${}^1$,\\ O.M. Teslyk${}^1$,  N.S. Yakovenko${}^1$,\\
A.O. Zaporozhchenko${}^1$,
E.V.~Gorbar${}^{1,2}$
\\
${}^1$ \it \small Faculty of Physics, Taras Shevchenko National University of Kyiv,\\
\it \small 64, Volodymyrs'ka str., Kyiv 01601, Ukraine\\
${}^2$ \it \small Bogolyubov Institute for Theoretical Physics, National Academy of Sciences of Ukraine,\\
\it \small 14-b, Metrolohichna str., Kyiv 03143, Ukraine
}
\date{}

\maketitle
\setcounter{equation}{0}
\setcounter{page}{1}%

\begin{abstract}
The dynamical friction force acting on a spatially extended probe (globular clusters and dwarf galaxies) moving in the environment of ultralight bosonic dark matter in the state of the Bose-Einstein condensate is determined. Modelling the probe as a Plummer sphere of radius $l_p$, the radial and tangential components of the dynamic friction force are found in an analytic form, which reduces in the limit $l_p \to 0$ to the corresponding analytic expressions obtained in the literature in the case of a point probe. The dependence of the dynamical friction force on boson particle mass
was analyzed and found to be non-monotonous in the interval $10^{-23} - 10^{-21}$eV. 

\end{abstract}

\section{Introduction}

Dynamical friction acting on objects moving through a galactic environment is an important and extensively studied phenomenon. It was first studied by Chandrasekhar \cite{Chandrasekhar} who analysed the gravitational drag on a moving star due to the fluctuating gravitational field of neighboring stars. Later this study was extended to the case of gaseous medium in \cite{Bondi,Dokuchaev,Ruderman,Rephaeli,Ostriker}.

Recently, models of ultralight dark matter (ULDM) with particle masses in the range 
$10^{-23}$ -- $10^{-21}$eV have attracted significant attention due to their intriguing phenomenology (for a review, see Refs.\cite{Chavanis:2015zua,Niemeyer:2019aqm,Hui:2021tkt,Ferreira}). These models successfully reproduce the large-scale structure of the Universe like cold dark matter (CDM) models and avoid some problems at the small scale faced by CDM. A distinctive feature of ULDM is the formation of the Bose-Einstein condensate (BEC) of ultralight bosons at galactic centers \cite{Salucci:2018hqu}.

As to the dynamical friction acting on moving objects in the ULDM environment, it has been studied for point probes in the case of fuzzy ULDM when the dark matter self-interaction is absent \cite{Hui:2016ltb,Lancaster_2020,Wang,Boey:2024dks}. Still, it has been found that even rather weak ULDM self-interaction can notably modify the frictional force affecting stellar motion \cite{Boudon:2022dxi,Hartman,Buehler:2022tmr,Berezhiani:2023vlo}. On larger scales, dynamical friction also plays an important role for motion and the evolution of more massive and spatially extended astrophysical objects such as globular clusters and dwarf galaxies. Since dynamical friction is proportional to the square of the moving object's mass, its influence on globular clusters can be considerably stronger compared to the case of individual stars \cite{Tremaine:1975}.

Globular clusters are often modeled (as well as sometimes dwarf galaxies) using the density profile of the Plummer sphere \cite{Plummer}. Since the de Broglie wavelength of dark matter particles composing ULDM can be significantly larger than the interstellar distances within a globular cluster, the dynamical friction effect is not simply the sum of individual stellar dynamical friction forces \cite{Lancaster_2020,Glennon}. We would like to mention also that the sizes of globular clusters (up to 10 pc) are much smaller than the de Broglie wavelength of the ULDM (approximately equal or larger 300 pc). Numerical studies have shown that for extended objects like globular clusters, the frictional force is weaker than for point probes of the same total mass, potentially alleviating the Fornax timing problem \cite{Blas,Hartman}.

In dwarf galaxies, like Fornax, globular clusters are inside the soliton core of ULDM in the BEC state. This motivates us to analyse in the present paper the dynamical friction force acting on a moving Plummer sphere in the environment of the BEC of ultralight bosonic dark matter.

Previously, the dynamical friction force acting on a circularly moving object (modelled as the Plummer sphere) in the BEC soliton core was determined in \cite{Gorkavenko:2024upe}. While the expression for the tangential force was obtained in an analytical form convenient for calculations, the radial component of the dynamical friction force was given as the Cauchy principal value of an integral over momentum and was computed only numerically. In this paper, we aim to integrate over momentum and obtain the radial component of the dynamical friction force in the same analytic form as it was derived for a point probe in \cite{Berezhiani:2023vlo}.

The paper is organized as follows. The dynamical friction for circularly moving Plummer sphere in the linear response approach is presented in Sec.\ref{sec:extended-body}. Analytic calculation of the imaginary and real components of this force is given in Sec.\ref{sec:analytic-calculation}. The dependence of dynamical friction force on boson particle mass $m$ is analyzed in Sec.\ref{Sec:4}. Conclusions are drawn in Sec.\ref{sec:Conclusion}.
	
\section{Dynamical friction force for circularly moving Plummer sphere}
\label{sec:extended-body}

In this section, we determine the dynamical friction force acting on a Plummer sphere, which moves on a circular orbit of radius $r_0$ in ultralight dark matter composed of ultralight bosonic particles of mass $m$ with constant angular velocity $\Omega$ in the steady-state regime.

Let us consider a Plummer sphere of radius $l_p$ and total mass $M$, whose mass density profile is given by
\begin{equation}
\rho_{Pl}(\mathbf{r})=\frac{3M}{4\pi l^3_p}\frac{1}{\left(1+\frac{r^2}{l^2_p}\right)^{5/2}}.
\label{Plummer-sphere}
\end{equation}
This density profile is approximately constant for $r<l_p$ and quickly decreases as $\sim 1/r^5$ for $r>l_p$. Using  
 \cite{Gradshteyn}, we easily find the following Fourier transform of the Plummer sphere mass density:
\begin{equation}
\rho_{Pl}(\mathbf{k})=Mkl_p\, K_1(kl_p)\equiv \rho_{Pl}(kl_p),
\label{density-momentum-1}
\end{equation}
where $K_1(x)$ is the modified Bessel function of the second kind. Since
$K_1(x) \to 1/x$ as $x \to 0$, we
find that $\rho_{Pl}(k l_p) \to M$ for $l_p \to 0$. As expected, this means that the mass density profile of Plummer sphere in momentum space tends to the mass density profile of a point probe given by $\rho_p(\mathbf{k})=M$.

To analyze the dynamical friction force, we follow the setup developed in \cite{Desjacques_2022,Buehler:2022tmr} and set the density of unperturbed ULDM to a constant value $\rho_0$. Then, moving Plummer sphere perturbs due to gravitational interaction the ULDM density $\rho_{DM}(t,\mathbf{r})=\rho_0(1+\alpha(t,\mathbf{r}))$.  Using the results of \cite{Desjacques_2022} for self-interacting ULDM and generalizing the corresponding analysis to the case under consideration, we obtain that the ULDM density inhomogeneity $\alpha(t,\mathbf{r})$ is governed by the following equation in the linear response approach:
\begin{equation}
\partial_t^2\alpha-c^2_s\nabla_{\mathbf{r}}^2\alpha+\frac{\nabla_{\mathbf{r}}^4\alpha}{4m^2}=4\pi G\rho_{Pl}(\mathbf{r}-\mathbf{r}_{CM}(t)).
\label{perturbation-sphere}
\end{equation}
Here, $c_s$ is the adiabatic sound velocity of DM superfluid
 which is given by the first derivative of pressure with respect to density \cite{Ferreira}
\begin{equation}\label{soundveloc}
 c_s=\frac{\sqrt{g\rho_{DM}}}{m},
\end{equation}
where $\rho_{DM}$ is the DM density, $g$ is the coupling constant of the DM self-interaction and $m$ is mass of the DM particle. 
Since pressure is proportional to $g$, sound velocity $c_s$ vanishes in fuzzy dark matter where $g=0$. Radius-vector 
$\mathbf{r}_{CM}(t)$ denotes the position of the center of mass of the moving Plummer sphere. The total dynamical friction force acting on the moving Plummer sphere (see, for more detailed consideration \cite{Gorkavenko:2024upe}) is given by
\begin{equation}
\mathbf{F}_{fr}(t)=(4\pi G)^2\rho_0\int\limits_{0}^{+\infty} d\tau\int \frac{d\omega d^3k}{(2\pi)^4}\,\frac{i\mathbf{k}}{\mathbf{k}^2} \frac{\rho_{Pl}(-\mathbf{k})\rho_{Pl}(\mathbf{k})\,e^{-i\omega\tau+i\mathbf{k}\mathbf{r}_{CM}(t)-i\mathbf{k}\mathbf{r}_{CM}(t-\tau)}}{-(\omega+i\epsilon)^2+c^2_s\mathbf{k}^2+\frac{\mathbf{k}^4}{4m^2}}.
\label{dynamical-force-sphere-total}
\end{equation}

Since $\rho_{Pl}(\mathbf{k})$ depends only on the absolute value of momentum $k$, the subsequent integration $d^3k=k^2dkd\Omega_k$ over $\Omega_k$ proceeds as in \cite{Berezhiani:2023vlo} and we obtain the following total dynamical friction force acting on the circularly moving Plummer sphere:
\begin{equation}
\mathbf{F}_{fr}(t)=-\frac{4\pi G^2M^2\rho_0}{c^2_s}\vec {\mathcal{F}},
\label{dynamical-force}
\end{equation}
where $\vec {\mathcal{F}}$ is dimensionless force whose nonzero radial and tangential components equal
\begin{equation}
\vec {\mathcal{F}}=\sum_{\ell=1}^{\ell_\text{\tiny max}}\sum_{m_l=-\ell}^{\ell-2}\gamma_{\ell m_l} \Bigg\{ \text{Re}\left(S_{\ell,\ell-1}^{m_l}-{S^{m_l+1}_{\ell,\ell-1}}^*\right)\hat{r} +  \text{Im}\left(S_{\ell,\ell-1}^{m_l}-{S^{m_l+1}_{\ell,\ell-1}}^*\right)\hat{\varphi}\Bigg\}.
\label{FDF1}
\end{equation}
Here
\begin{multline}
    \gamma_{\ell m_l} = (-1)^{m_l} \frac{(\ell-m_l)!}{(\ell-m_l-2)!} \times \\ \times \Bigg\{ \Gamma\left(\frac{1-\ell-m_l}{2}\right) \Gamma\left(1+\frac{\ell-m_l}{2}\right)
    \Gamma\left(\frac{3-\ell+m_l}{2}\right)\Gamma\left(1+\frac{\ell+m_l}{2}\right) \Bigg\}^{-1}
\end{multline}
and the key quantity which defines the dynamical friction force is
\begin{equation}
S^{m_l}_{\ell,\ell-1} =\frac{c^2_s}{M^2}\int\limits^{+\infty}_0 \frac{kdk\,\rho^2_{Pl}(k l_p)\,j_{\ell}(kr_0)j_{\ell-1}(kr_0)}{c^2_sk^2+\frac{k^4}{4m^2}-(m_l\Omega+i\epsilon)^2}, \\
\label{S-function-1}
\end{equation}
where $\epsilon \to +0$, $\ell$ and $m_l$ are the azimuthal and quantum numbers, respectively, $j_{\ell}(x)$ is the spherical Bessel function. In the case of a point probe where $\rho_{Pl}(k l_p) \to M$, the integral over $k$ in Eq.(\ref{S-function-1}) was calculated analytically in \cite{Berezhiani:2023vlo}.
In the next section, we calculate this integral for $\rho_{Pl}(k l_p)$ and find the analytic expressions for the real and imaginary parts of $S^{m_l}_{\ell,\ell-1}$.

\section{Analytic calculation of $S_{\ell,\ell_1}^{m_l}$}
\label{sec:analytic-calculation}

Representing the function $\rho_{Pl}(k l_p)$ in the integral form
\begin{equation}
    \rho_{Pl}(k l_p)=M\int\limits_0^{\infty}\frac{\cos(kl_px)}{(1+x^2)^{3/2}}\,dx
\end{equation}
and taking into account that the integrand is an even function of $x$, we obtain
\begin{equation}
    \rho_{Pl}^2(k l_p)=\frac{M^2}{4}\int\limits_{0}^{\infty}d\,u \int\limits_{0}^{\infty}d\,v  \frac{e^{i l_p u k}+e^{-i l_p u k}}{(1+\frac14(u+v)^2)^{3/2}(1+\frac14(u-v)^2)^{3/2}}
    = \frac{M^2}{4}\eta(k l_p),
\label{density-square}
\end{equation}
where $u=x+y$, $v=x-y$, and
\begin{align}
    &\eta(k l_p) = \int\limits_{0}^{\infty} \int\limits_{0}^{\infty}\frac{e^{i l_p u k}+e^{-i l_p u k}}{f(u,v)}d\,ud\,v,
    \label{eta} \\
    &f(u,v)=\left(1+\frac14(u+v)^2\right)^{3/2}\left(1+\frac14(u-v)^2\right)^{3/2}.
\end{align}

The spherical Bessel functions $j_l(x)$ can be expressed through the spherical Hankel functions $h^{(1,2)}_l(x)$
\begin{equation}
    j_l(x)=\frac{1}{2}(h_l^{(1)}(x)+h_l^{(2)}(x)).
\end{equation}

Further, it is convenient to split the product $j_l(x)j_{l-1}(x)$ into two components, which include as a factor only one exponential function $e^{2ix}$ or $e^{-2ix}$, see for details, e.g., \cite{Gradshteyn},
\begin{multline}
  4j_l(x)j_{l-1}(x)= \underbrace{h_l^{(1)}(x)h_{l-1}^{(1)}(x)+h_l^{(1)}(x)h_{l-1}^{(2)}(x)}_{g_1(x)} + \\
\underbrace{h_l^{(2)}(x)h_{l-1}^{(2)}(x)+h_l^{(2)}(x)h_{l-1}^{(1)}(x)}_{g_2(x)} = g_1(x)+g_2(x),  
\end{multline}
where
\begin{align}
    &g_1(x)=a_1(x)+ib_1(x)+(a_2(x)-ib_2(x))e^{2ix},  \nonumber \\
    &g_2(x)=a_1(x)-ib_1(x)+(a_2(x)+ib_2(x))e^{-2ix}, \nonumber \\
    &g_1(-x)=-g_2(x). 
\end{align}

Here $a_i(x)$, $b_i(x)$ are polynomials in inverse powers of $x$. Then Eq.(\ref{S-function-1}) takes the form
\begin{equation}
    S_{l,l-1}^{m_l}\!=\!\frac{c^2_sm^2}{8}\!\!\int\limits_{-\infty}^\infty\!\frac{k\eta(k l_p)}{\Pi(k^2)}\,(g_1(kr_0)+g_2(kr_0))dk,    
    \label{S-coefficient}
\end{equation}
where $\Pi(k^2)=(k^2+\varkappa^2)(k^2-k_3^2)$ and we used
\begin{equation}
    4m^2c^2_sk^2+k^4-4m^2m^2_l\Omega^2 
    = (k+i\varkappa)(k-i\varkappa)(k-k_3)(k+k_3)
\end{equation}
with
\begin{equation}
    \varkappa=\sqrt{2}mc_s\left(\sqrt{1+\frac{m_l^2\Omega^2}{m^2c^4_s}}+1\right)^{1/2}, \quad
    k_{3}=\sqrt{2}mc_s\left(\sqrt{1+\frac{m_l^2\Omega^2}{m^2c^4_s}}-1\right)^{1/2}.
\label{poles}
\end{equation}
The function $\eta(kl_p)$ is defined in Eq.(\ref{density-square}) and has the structure 
$$\eta(k  l_p)=\int\limits_{0}^{\infty} \int\limits_{0}^{\infty}\frac{d\,ud\,v}{f(u,v)}(e^{i l_p u k}+e^{-i l_p u k}).$$
Note that the functions $kg_1(kr_0)$ and $kg_2(kr_0)$ have only one simple pole at $k=0$. Hence, we can split the integral in Eq.(\ref{S-coefficient}) into two integrals
\begin{equation}
    S_{l,l-1}^{m_l}=\frac{c^2_sm^2}{8}\hspace{0.3em}-\hspace{-1.35em}\int\limits_{-\infty}^\infty\frac{k\eta(k  l_p)}{\Pi(k^2)}\,g_1(kr_0)dk 
    + \frac{c^2_sm^2}{8}\hspace{0.3em}-\hspace{-1.35em}\int\limits_{-\infty}^\infty\frac{k\eta(k  l_p)}{\Pi(k^2)}\,g_2(kr_0)dk=S_1+S_2,
\end{equation}
which converge in the sense of the Cauchy principal value (which is denoted by  $\int\hspace{-0.9em}-$). 

By replacing $k\to-k$ in the second integral $S_2$ and taking into account that $g_2(-k)=-g_1(k)$, we find that $S_2=S_1$. Therefore, 
\begin{equation}
    S_{l,l-1}^{m_l}=2S_1=\frac{c^2_sm^2}{4}\int\limits_{0}^\infty \int\limits_{0}^\infty \frac{du\,dv}{f(u,v)} 
     \underbrace{\hspace{0.3em}-\hspace{-1.3em}\int\limits_{-\infty}^\infty\frac{e^{i l_p u k}+e^{-il_p u k}}{\Pi(k^2)}kg_1(kr_0)\,dk}_{J}.
\end{equation}
For $J$, we have
\begin{equation}
   J=\int\limits_C \frac{e^{i l_p u k}+e^{-i l_p u k}}{\Pi(k^2)}kg_1(kr_0)\,dk 
   + i\pi \mathop{\mathrm{res}}\limits_{k=0} \left( \frac{e^{i l_p u k}+e^{-i l_p u k}}{\Pi(k^2)}kg_1(kr_0)\right),
\end{equation}
where the contour $C$ corresponds to the integration from $-\infty$ to $+\infty$ along the real axis $k$ and 
passing around the point $k=0$ in the upper half-plane.  Further,
\begin{multline}
    \int\limits_C\frac{e^{i l_p u k}+e^{-i l_p u k}}{\Pi(k^2)}kg_1(kr_0)\,dk = \\
    = \int\limits_C\frac{e^{i l_p u k}}{\Pi(k^2)}kg_1(kr_0)\,dk+\int\limits_C\frac{e^{-i l_p u k}}{\Pi(k^2)}kg_1(kr_0)\,dk 
    = J_1+J_2.
\end{multline}

\begin{figure*}[t]
    \centering
    \includegraphics[width=0.70\textwidth]{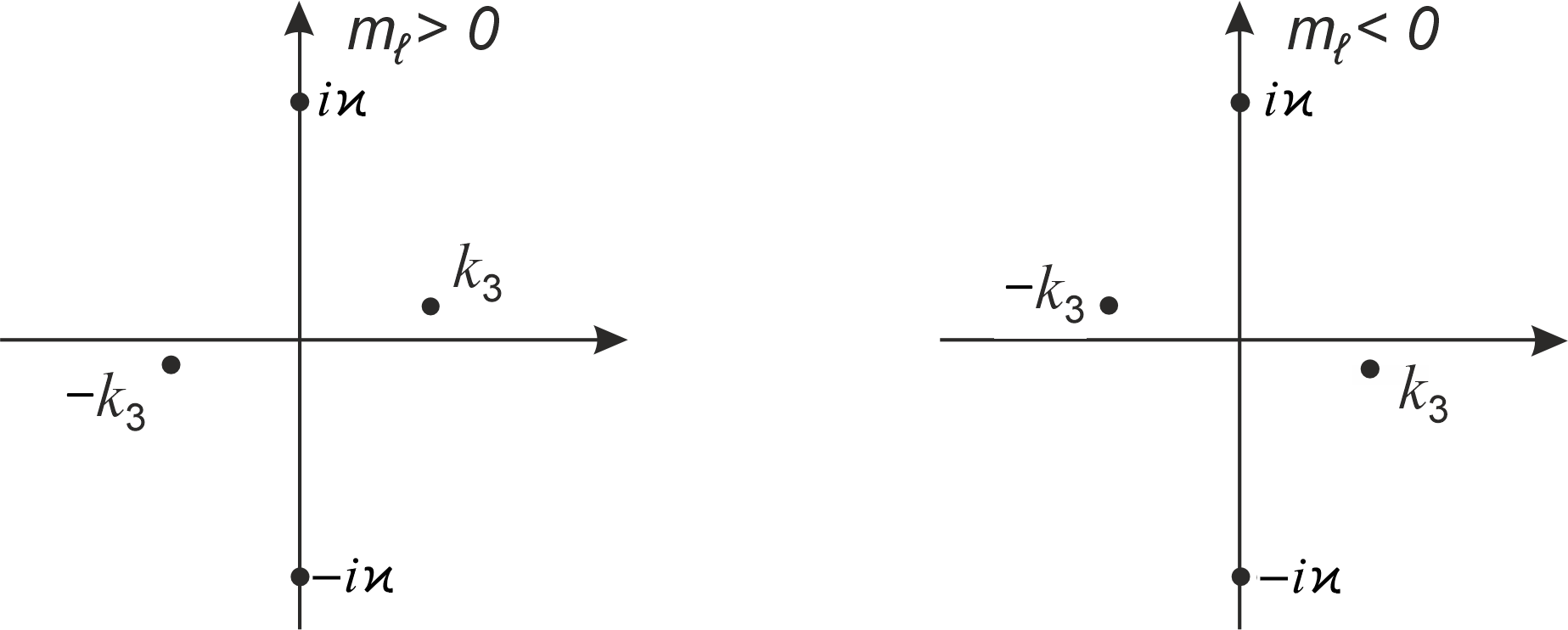}
    \caption{Poles of the integrand in $J$ in the complex plane $k$ for $m_l>0$ (left panel) and $m_l<0$ (right panel).}
    \label{fig:KorshProf}
\end{figure*}

To calculate the integral $J_1$, we close the contour in the upper half-plane and apply the Cauchy formula. We begin with the case $m_l>0$. In the upper half-plane, we have two simple poles at $k=k_3$ and $k=i\varkappa$, see Fig.~\ref{fig:KorshProf}. We find 
\begin{equation}
J_1=
\frac{\pi i}{\varkappa^2+k_3^2}\Big(
-e^{-\varkappa l_p u}g_1(i\varkappa r_0)+e^{i l_p u k_3}g_1(k_3 r_0)\Big).
\end{equation}
To calculate $J_2$, we should consider two cases with $l_p u>2r_0$ and $l_p u<2r_0$. We have
\begin{equation}
J_2=\int\limits_C\frac{ke^{-i l_p u k}g_1(kr_0)}{\Pi(k^2)}\,dk
= \int\limits_C\frac{ke^{-i l_p u k}}{\Pi(k^2)}(\underbrace{a_1+ib_1}_{h_l^{(1)}h_{l-1}^{(2)}}+ (\underbrace{a_2-ib_2)e^{2ikr_0}}_{h_l^{(1)}h_{l-1}^{(1)}})\,dk.
\end{equation}

Case $l_p u>2r_0$. Closing the contour in the lower half-plane and applying the Cauchy formula, we find
\begin{equation}
J_2=\frac{\pi i}{\varkappa^2+k_3^2}\left(
-e^{-\varkappa l_p u }g_2(i\varkappa r_0)+e^{i l_p u k_3}g_2(k_3 r_0)\right) 
- 2\pi i \mathop{\mathrm{res}}\limits_{k=0} \left( \frac{kg_1(kr_0)e^{-i l_p u k}}{\Pi(k^2)} \right),
\end{equation}
where the last term is due to the pole at $k=0$ and we took into account that $g_1(-z)=-g_2(z)$.

Case $ l_p u <2r_0$.  Now for the part $h_l^{(1)}h_{l-1}^{(2)}$ we need to close the contour in the lower half-plane, and for the part with $h_l^{(1)}h_{l-1}^{(1)}$ we need to close the contour in the upper half-plane. We obtain
\begin{multline}
    J_2=\int\limits_C\frac{ke^{-i l_p u k}}{\Pi(k^2)} 
    \left[h_l^{(1)}(kr_0)h_{l-1}^{(2)}(kr_0)+h_l^{(1)}(kr_0)h_{l-1}^{(1)}(kr_0)\right]\,dk =\\
    = \frac{\pi i}{\varkappa^2+k_3^2}\Big(-e^{-\varkappa l_p u}h_l^{(2)}(i\varkappa r_0)h_{l-1}^{(1)}(i\varkappa r_0)
    +e^{il_p u k_3}h_l^{(2)}(k_3r_0)h_{l-1}^{(1)}(k_3r_0)\Big) - \\
    - 2\pi i \mathop{\mathrm{res}}\limits_{k=0} \left( \frac{k\,h_l^{(1)}(kr_0)h_{l-1}^{(2)}(kr_0)e^{-il_p u k}}{\Pi(k^2)}\right) + \\
    + \frac{\pi i}{\varkappa^2+k_3^2}\Big(-e^{\varkappa l_p u}h_l^{(1)}(i\varkappa r_0)h_{l-1}^{(1)}(i\varkappa r_0) 
    + e^{-i l_p u k_3}h_l^{(1)}(k_3r_0)h_{l-1}^{(1)}(k_3r_0)\Big).
\end{multline}
Thus, we have the following results for $J=J_1+J_2+i\pi \mathop{\mathrm{res}}\limits_{k=0} \left( \frac{e^{i l_p u k}+e^{-i l_p u k}}{\Pi(k^2)}kg_1(kr_0) \right)$.\vspace{0.5em}

(i) Case $l_p u>2r_0$. The integral $J$ equals
\begin{equation}
J=\frac{4\pi i}{\varkappa^2+k_3^2}\Big(
-e^{-\varkappa l_p u}j_l(i\varkappa r_0)j_{l-1}(i\varkappa r_0) 
+ e^{i l_p u k_3}j_l(k_3r_0)j_{l-1}(k_3r_0)\Big).
\end{equation}
The residue term
\begin{equation}
-2\pi \mathop{\mathrm{res}}\limits_{k=0} \left( \frac{kg_1(kr_0)\sin (l_p u k)}{\Pi(k^2)} \right)
\end{equation}
vanishes because numerator is regular at $k=0$.

(ii) Case $l_p u<2r_0$. We have
\begin{multline}\label{res}
    J=\frac{i\pi}{\varkappa^2+k_3^2}\Big(
-e^{-\varkappa l_p u}\left[h_l^{(1)}(i\varkappa r_0)h_{l-1}^{(1)}(i\varkappa r_0) 
+ h_l^{(1)}(i\varkappa r_0)h_{l-1}^{(2)}(i\varkappa r_0)+ h_l^{(2)}(i\varkappa r_0)h_{l-1}^{(1)}(i\varkappa r_0) \right] - \\
- e^{\varkappa l_p u}h_l^{(1)}(i\varkappa r_0)h_{l-1}^{(1)}(i\varkappa r_0) 
+ e^{-i l_p u k_3}h_l^{(1)}(k_3r_0)h_{l-1}^{(1)}(k_3r_0) + \\
+ e^{i l_p u k_3}\left[h_l^{(1)}(k_3r_0)h_{l-1}^{(1)}\!(k_3r_0)+ h_l^{(1)}\!(k_3r_0)h_{l-1}^{(2)}(k_3r_0) 
+ h_l^{(2)}(k_3r_0)h_{l-1}^{(1)}(k_3r_0)\right]\Big) + \\
+ 2\pi i \mathop{\mathrm{res}}\limits_{k=0} \Bigg( \frac{kh_l^{(1)}(kr_0)}{\Pi(k^2)}\big(e^{i l_p u k}j_{l-1}(kr_0) 
+ ie^{-i l_p u k}y_{l-1}(kr_0)\big)\Bigg),
\end{multline}
where $y_{l}(x)=(h_l^{(1)}(x)-ih_l^{(2)}(x))/2i$ is the spherical Neumann function.

For the imaginary part of $J$,
we obtain \vspace{-0.5em}
\begin{equation}
\mathrm{Im}\,J = \frac{4\pi}{\varkappa^2+k_3^2}j_l(k_3r_0)j_{l-1}(k_3r_0)\cos(l_p u k_3), 
\end{equation}
where  $l_p u>2r_0$ and
\begin{equation}
\mathrm{Im}\,J  = \frac{4\pi}{\varkappa^2+k_3^2}j_l(k_3r_0)j_{l-1}(k_3r_0)\cos(l_p u k_3) 
- 2\pi \mathop{\mathrm{res}}\limits_{k=0} \left( \frac{k\cos(l_p u k)}{\Pi(k^2)}y_ly_{l-1}(kr_0) \right), 
\label{imj}
\end{equation}
for $l_p u<2r_0$.  
The residue term in Eq.(\ref{imj}) is zero as it is the residue of an even function of $k$. Therefore, for the imaginary part of $S_{l,l-1}^{m_l}$, we find the following expression:\vspace{-0.5em}
\begin{multline}
  \mathrm{Im}\,S_{l,l-1}^{m_l}=\frac{\pi c_s^2m^2j_l(k_3r_0)j_{l-1}(k_3r_0)}{\varkappa^2+k_3^2} 
  \int\limits_0^{\infty}\!\!\int\limits_0^{\infty}\frac{\cos(l_p u k)}{f(u,v)}dudv = \\
  = \frac{2\pi c_s^2m^2\rho^2_{Pl}(k_3  l_p)j_l(k_3r_0)j_{l-1}(k_3r_0)}{M^2(\varkappa^2+k_3^2)},
\end{multline}
which exactly coincides with the result previously obtained in \cite{Gorkavenko:2024upe}. In addition, taking the limit of vanishing Plummer sphere radius ($l_p \to 0$) and using that $\rho_{Pl}(x) \to M$ as $x\to 0$, this expression completely agrees with the imaginary part found in \cite{Berezhiani:2023vlo} in the case of a point probe.

Let us proceed to the real part of $S_{l,l-1}^{m_l}$ and begin with the real part of $J$. We have
\vspace{-0.5em}
\begin{equation}
    \mathop{\Re\mathrm{e}} J=-\frac{4\pi}{\varkappa^2+k_3^2}\Big(
ie^{-\varkappa l_p u}j_l(i\varkappa r_0)j_{l-1}(i\varkappa r_0) 
+j_l(k_3r_0)j_{l-1}(k_3r_0)\sin(l_p u k_3)\Big)
\label{real-final-greater}
\end{equation}
for $l_p u>2r_0$ and\vspace{-1em}
\begin{multline}
  \mathop{\Re\mathrm{e}} J=\frac{\pi}{\varkappa^2+k_3^2}\Big(
-ie^{-\varkappa l_p u}\left[2j_l(i\varkappa r_0)h_{l-1}^{(1)}(i\varkappa r_0) 
+ h_l^{(1)}(i\varkappa r_0)h_{l-1}^{(2)}(i\varkappa r_0)\right] -\\- e^{\varkappa l_p u}h_l^{(1)}(i\varkappa r_0)h_{l-1}^{(1)}(i\varkappa r_0) 
- \sin(l_p u k_3)\left[h_l^{(1)}(k_3r_0)h_{l-1}^{(2)}(k_3r_0) 
+ h_l^{(2)}(k_3r_0)h_{l-1}^{(1)}(k_3r_0)\right] - \\
- 2\cos(l_p u k_3)\left[\vphantom{Y^{\frac12}}j_l(k_3r_0)y_{l-1}(k_3r_0) 
+ y_l(k_3r_0)j_{l-1}(k_3r_0)\right]\Big)-\\ - \frac{2\pi}{\varkappa^2k_3^2r_0^2} 
- 2\pi \mathop{\mathrm{res}}\limits_{k=0} \left( \frac{k}{\Pi(k^2)}y_l(kr_0)y_{l-1}(kr_0)\sin(l_p u k) \right)
\label{real-final-smaller}
\end{multline}
for $l_p u<2r_0$. The last two terms in (\ref{real-final-smaller}) are due to the residue term in expression (\ref{res})
\begin{multline}
    i \mathop{\mathrm{res}}\limits_{k=0} \left( \frac{kh_l^{(1)}(kr_0)}{\Pi(k^2)}\big(e^{i l_p u k}j_{l-1}(kr_0)+ie^{-i l_p u k}y_{l-1}(kr_0)\big) \right) 
    =\\= -\mathrm{res}\hspace{-0.55cm}\raisebox{-0.25cm}{$\scriptstyle k=0$}\,\Big(\frac{k}{\Pi(k^2)}\big(\sin(l_p u k)j_l(kr_0)j_{l-1}(kr_0) 
   + \cos(l_p u k)j_l(kr_0)y_{l-1}(kr_0) +\\
    + \sin(l_p u k)y_l(kr_0)y_{l-1}(kr_0 )
    + \cos(l_p u k)y_l(kr_0)j_{l-1} (kr_0)\big)\Big),
\label{residue}
\end{multline}
where we took into account that the residue of an even function at the zero value of its argument vanishes. The first two terms are regular at zero, and therefore, their contribution is zero. The last term has a simple pole at zero and can be easily calculated using the asymptotics
\begin{align}
& j_{l-1}(x)=\frac{x^{l-1}}{(2l-1)!!}+O(x^{l+1}), \\
& y_{l}(x)=-\frac{(2l-1)!!}{x^{l+1}}+O(x^{l-1}).
\end{align}

We find
\begin{equation}
\mathop{\mathrm{res}}\limits_{k=0} \left( \frac{\cos(l_p u k)}{\Pi(k^2)}ky_l(kr_0)j_{l-1}(kr_0) \right)=\frac{1}{\varkappa^2k_3^2r_0^2}.
\end{equation}
As to the residue term in Eq.(\ref{real-final-smaller}), it equals
\begin{multline}
\Delta(u)= \mathop{\mathrm{res}}\limits_{k=0} \left( \frac{k\sin(l_p u k)}{\Pi(k^2)}y_l(kr_0)y_{l-1}(kr_0) \right) =\\
-\frac{1}{\varkappa^2+k_3^2} \sum_{n=0}^{l-1}\frac{(2n)!(l+n)!}{4^n(n!)^2(l-n-1)!} 
\times \sum_{m=0}^{n} \Big(\frac{1}{k_3^{2m+2}}+\frac{(-1)^m}{\varkappa^{2m+2}}\Big)\frac{(-1)^{n-m}(l_p u)^{2(n-m)+1}}{r_0^{2n+3}(2n-2m+1)!}
\label{residue-Delta}
\end{multline}
with details of its calculation given in Appendix.

Finally, using Eqs.(\ref{real-final-greater}), (\ref{real-final-smaller}), and (\ref{residue}), we obtain the following real part of $S_{l,l-1}^{m_l}$ for $m_l>0$:\vspace{-0.5em}
\begin{multline}\label{ReS}
    \mathop{\Re\mathrm{e}} S_{l,l-1}^{m_l}=\frac{\pi c_s^2m^2}{\varkappa^2+k_3^2}\Big(-iR_1j_l(i\varkappa r_0)j_{l-1}(i\varkappa r_0) 
    - R_2j_l(k_3 r_0)j_{l-1}(k_3 r_0) - \frac{Q_2}{2}(y_l(k_3r_0)y_{l-1}(k_3r_0) + \\ +j_l(k_3r_0)j_{l-1}(k_3r_0)) - \frac{Q_3}{2}(j_l(k_3r_0)y_{l-1}(k_3r_0) 
    + y_l(k_3r_0)j_{l-1}(k_3r_0))-\\- \frac{i Q_1^{-}}{4}(2j_l(i\varkappa r_0)h_{l-1}^{(1)}(i\varkappa r_0) 
    + h_l^{(1)}(i\varkappa r_0)h_{l-1}^{(2)}(i\varkappa r_0)) -\frac{i Q_1^{+}}{4}h_l^{(1)}(i\varkappa r_0)h_{l-1}^{(1)}(i\varkappa r_0)\Big)\\-\frac{\pi c_s^2m^2}{2\varkappa^2k_3^2r_0^2}Q_4-\frac{\pi m^2c_s^2}{2}Q_5,
\end{multline}
where
\begin{align}
& R_1\!=\!\int\limits_{\frac{2r_0}{l_p}}^\infty du\!\int\limits_{0}^\infty\frac{e^{-\varkappa l_pu}}{f(u,v)}\,dv,\quad R_2\!=\!\int\limits_{\frac{2r_0}{l_p}}^\infty du\!\int\limits_{0}^\infty \frac{\sin(k_3l_pu)}{f(u,v)}\,dv, \\
& Q_1^{\pm}\!\!=\!\!\int\limits_{0}^{\frac{2r_0}{l_p}} du \!\int\limits_{0}^\infty\frac{e^{\pm\varkappa l_pu}}{f(u,v)}\,dv,\quad Q_2\!\!=\!\!\int\limits_{0}^{\frac{2r_0}{l_p}} du\!\int\limits_{0}^\infty \frac{\sin(k_3l_pu)}{f(u,v)}\,dv, \\
& Q_3\!=\!\int\limits_{0}^{\frac{2r_0}{l_p}} du\int\limits_{0}^\infty \frac{\cos(k_3l_pu)}{f(u,v)}\,dv,\quad Q_4\!=\!\int\limits_{0}^{\frac{2r_0}{l_p}}du\int\limits_{0}^\infty \frac{dv}{f(u,v)}, \\
& Q_5=\int\limits_{0}^{\frac{2r_0}{l_p}}du\int\limits_{0}^\infty \frac{\Delta(u) dv}{f(u,v)}.
\end{align}
In the case $m_l<0$, we should replace $k_3\to-k_3$. Since expression (\ref{ReS}) is an even function of $k_3$, the same expression for the real part of $S_{l,l-1}^{m_l}$ still applies for $m_l<0$.

It should be noted that the real part $\mathop{\Re\mathrm{e}} S_{l,l-1}^{m_l}$ was previously calculated and analyzed numerically in Ref.\cite{Gorkavenko:2024upe}. It is given as the Cauchy principal value of Eq.(\ref{S-function-1}), i.e.,
\begin{equation}
\mathop{\Re\mathrm{e}}\, S^{m_l}_{\ell,\ell-1}=
\frac{4 m^2 c^2_s r_0^2}{ \hbar^2M^2}  \hspace{0.3em}-\hspace{-1.33em}\int\limits_0^{+\infty} \frac{x dx\,\rho^2_{Pl}(x l_p/r_0)\,j_{\ell}(x)j_{\ell-1}(x)}{x^4+\frac{4m^2c^2_sr_0^2}{\hbar^2}\, x^2-4 m_l^2\,\frac{ m^2\Omega^2r_0^4}{\hbar^2} }.
\label{S-function-1vp}
\end{equation}
We checked numerically that Eqs.\eqref{ReS} and \eqref{S-function-1vp} yield identical results, i.e., they present the same quantity in different forms. Finally, one can easily verify that the real part $\mathop{\Re\mathrm{e}} S_{l,l-1}^{m_l}$ in the limit $l_p\rightarrow 0$ coincides with the real part for a point probe found in \cite{Berezhiani:2023vlo}.

Using the obtained analytical formulas for the dynamical friction force, we analyse in the next section the dependence of the dynamical friction force on boson particle mass $m$.

\section{Dependence of dynamical friction force on boson particle mass}
\label{Sec:4}

The radial and tangential components of the dimensionless dynamical friction force $\vec {\mathcal{F}}$ given by Eq.\eqref{FDF1} acting on the circularly moving Plummer sphere were determined numerically in \cite{Gorkavenko:2024upe} and plotted there as functions of the Mach number ${\cal M}$ and dimensionless orbital radius $a =m c_sr_0/ \hbar$ for different values of $l_p/r_0$.
The Mach number is defined as $v/c_s$, where $v$ is the absolute value of the velocity of the probe in ULDM and {$c_s$ is the adiabatic sound speed} of DM superfluid (see for details \cite{Gorkavenko_2024}).

\begin{figure*}[t]
    \centering
    \includegraphics[width=\textwidth]{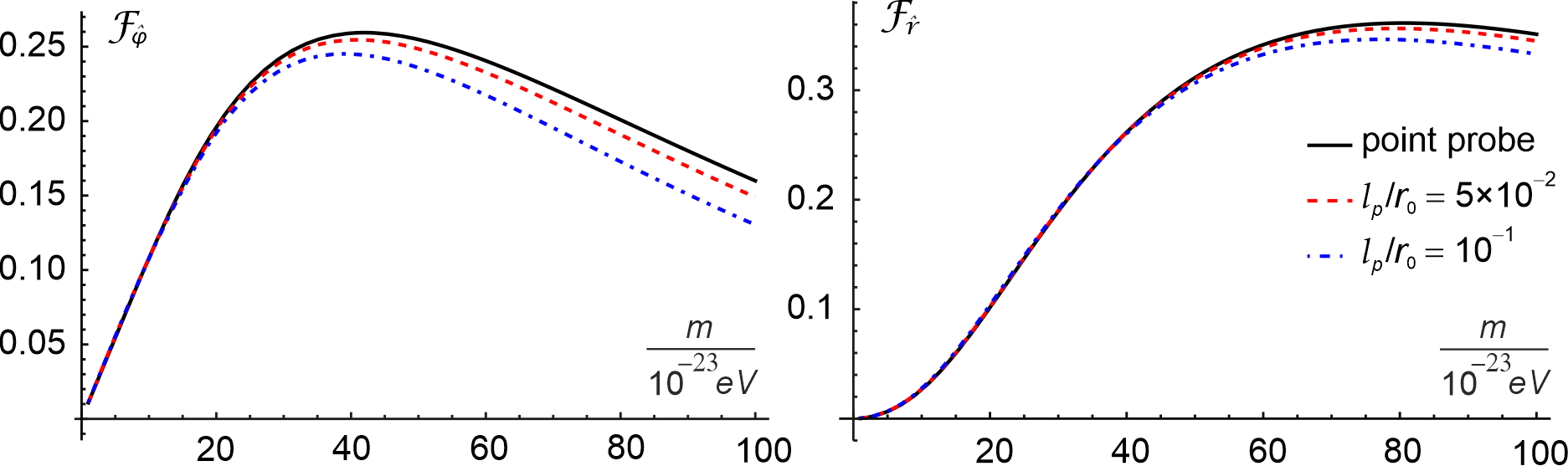}
    \caption{Tangential (left panel) and radial (right panel) components of the dimensionless dynamical friction force $\vec {\mathcal{F}}$ given by Eq.\eqref{FDF1} as a function of the mass of the dark matter particle $m$ in the interval $10^{-23} - 10^{-21}$eV at fixed orbital radius $r_0$ for a point probe and the Plummer sphere for a typical globular cluster in the Fornax dwarf galaxy.}
    \label{fig:mass}
\end{figure*}

Since the mass of the DM particle $m$ is not fixed in the ULDM model, we plot in Fig.\ref{fig:mass} the dependence of dimensionless dynamical friction force on the mass parameter $m$ in the interval $10^{-23} - 10^{-21}$eV for a typical globular cluster in the Fornax dwarf galaxy \cite{Lancaster_2020} with orbital radius $r_0=668$ pc and density of dark matter at this radius $\rho_{DM}=2\cdot 10^7$ M${}_{\odot}$/kpc${}^3$.

As one can see, the dependence of both radial and tangential components is not monotonic. The dynamical friction force grows with $m$, attains a maximum, and then decreases. While both components weakly depend on $m$  for $m \gtrsim 2\cdot10^{-22}$eV, they notably decrease as $m$ tends to $10^{-23}$eV. Obviously, the dynamical friction force for the Plummer sphere differs more strongly from that for a point probe of the same mass for larger values of the ratio $l_p/r_0$ and $m$.

\section{Conclusions}
\label{sec:Conclusion}

We studied the force of dynamic friction acting on a spatially extended probe (globular clusters and dwarf galaxies) moving in galactic ultralight dark matter in the state of the Bose-Einstein condensate.
These objects are often modelled as Plummer spheres. The dynamical friction force for a Plummer sphere moving in ultralight dark matter was previously considered in \cite{Gorkavenko:2024upe}.   
 Numerically studying the radial and tangential components of the dynamical friction force, it was found that the dynamical friction force for the Plummer sphere
deviates from that for a point probe of the same mass for a sufficiently large ratio
of the Plummer sphere radius to its orbital radius, as well as for large values of the
Mach number. This study confirmed the relevance of finite-size effects for the dynamical friction force in the case of globular clusters and dwarf galaxies.

The radial component of the dynamical friction force was given in  \cite{Gorkavenko:2024upe} as the Cauchy principal value of an integral over momentum and was computed only numerically. In this paper integrating over momentum, we determined the radial component of the dynamical friction force in the same form as for a point probe obtained in  \cite{Berezhiani:2023vlo}. We plotted in Fig.\ref{fig:mass} the dependence of the tangential and radial components of dimensionless dynamical friction force on the boson particle mass $m$ in the interval $10^{-23} - 10^{-21}$eV for a typical globular cluster in the Fornax dwarf galaxy.
We found that the dependence of both radial and tangential components is not monotonic. In addition, the dynamical friction force for the Plummer sphere differs more strongly from that for a point probe for larger values of the ratio $l_p/r_0$ and $m$.

We checked that our analytical expressions for the radial and tangential components of dynamical friction force reproduce in the limit of vanishing Plummer sphere radius ($l_p\rightarrow 0$) the corresponding expressions derived in  \cite{Berezhiani:2023vlo}. Comparing our results numerically with those in \cite{Gorkavenko:2024upe} shows their complete agreement with each other. Since the expressions for the dynamic friction acting on a finite-sized body are quite complex and their calculation requires numerical methods and a large amount of machine time, we think that the analytic expressions for the dynamic friction force obtained in this paper can be useful for practical calculations. In particular, they can be used
in future studies to improve our understanding of the ULDM impact on the orbital dynamics of extended astrophysical systems.

We would like to mention also that, unlike dwarf galaxies, globular clusters in the Milky Way are typically situated at distances larger than the central core of radius around 1 kpc in the state of
the Bose-Einstein condensate of ultralight bosons. At such distances, quantum wave interference gives rise to stochastically distributed de-Broglie-scale granulation \cite{Schive:2014dra,Schwabe:2016rze,Mocz:2017wlg,Veltmaat:2018dfz}, which is phenomenologically described via a dissipative term in the Gross-Pitaevskii equation \cite{Chavanis:2018pkx}, and we plan to study its role in the future.

\vspace{5mm}

\centerline{\bf Acknowledgements}
\vspace{5mm}

The work of E.V.G., T.V.G, V.M.G, and A.V.Z. was partially supported by the project 'Search for dark matter and particles beyond the Standard Model' of the Ministry of Education and Science of Ukraine (25BF051-01). The authors are grateful
to A.I. Yakimenko for fruitful
discussions and helpful comments.

\newpage

{ \setcounter{equation}{0}
\renewcommand{\theequation}{A.\arabic{equation}}
\section*{Appendix}

The residue 
\begin{equation}\label{res1}
\Delta(u)= \mathop{\mathrm{res}}\limits_{k=0} \left( \frac{k\sin(l_p u k)}{\Pi(k^2)}y_l(kr_0)y_{l-1}(kr_0) \right)
\end{equation}
not change if we replace $y_ly_{l-1}$ by $j_lj_{l-1}+y_ly_{l-1}$, i.e.,
\begin{equation}
    \Delta(u) = \mathop{\mathrm{res}}\limits_{k=0} \Bigg( \frac{k\sin(l_p u k)}{\Pi(k^2)}(j_l(kr_0)j_{l-1}(kr_0) 
    + y_l(kr_0)y_{l-1}(kr_0)) \Bigg)
\end{equation}
because $j_l(kr_0)j_{l-1}(kr_0)$ is regular at $k=0$. We introduce 
\begin{equation}
  Q_l(x)=j_l(x)j_{l-1}(x)+y_l(x)y_{l-1}(x)  
\end{equation}
because unlike $y_l(x)y_{l-1}(x)$ the function $Q(x)$ is a polynomial in inverse powers of $x$.

Definitions
\begin{align}
& j_l(x)=x^l\left(-\frac{1}{x}\frac{d}{dx}\right)^l\frac{\sin x}{x}, \\
& y_l(x)=-x^l\left(-\frac{1}{x}\frac{d}{dx}\right)^l\frac{\cos x}{x}
\end{align}
imply that
\begin{equation}
j_l(x)=-j_{l-1}'(x)+(l-1)\frac{j_{l-1}(x)}{x}
\end{equation}
and a similar relation holds for $y_l(x)$. Therefore,
\begin{multline}
    Q_l(x)=-\frac{1}{2}\frac{d}{dx}(j_{l-1}^2(x)+y_{l-1}^2(x)) 
    +\frac{l\!-\!1}{x}(j_{l-1}^2(x)+y_{l-1}^2(x))= \\
    = \left(-\frac{1}{2}\frac{d}{dx}+\frac{l\!-\!1}{x}\right) (j_{l-1}^2(x)+y_{l-1}^2(x)).
\end{multline}
Further, we have for $j_l^2(x)+y_l^2(x)$ \cite{Frank}
\begin{equation}\label{f1}
j_l^2(x)+y_l^2(x)=\sum_{n=0}^l\frac{(2n)!(n+l)!}{4^n(n!)^2(l-n)!}\frac{1}{x^{2n+2}}.
\end{equation}
Then using (\ref{f1}), we find
\begin{equation}
    Q_l(x)=\sum_{n=0}^{l-1}\frac{(2n)!(n+l)!}{4^n(n!)^2(l-n-1)!}\frac{1}{x^{2n+3}}.
\end{equation}
In view of the Taylor expansions
\begin{multline}
    \frac{1}{\Pi(k^2)} = \frac{1}{(k^2 + \varkappa^2)(k^2 - k_3^2)} 
    = \frac{1}{\varkappa^2 + k_3^2}\left(\frac{1}{k^2 - k_3^2}-\frac{1}{k^2 + \varkappa^2}\right) = \\
    = -\frac{1}{\varkappa^2 + k_3^2} \sum_{m=0}^{\infty} \left(\frac{1}{k_3^{2m+2}}+\frac{(-1)^m}{\varkappa^{2m+2}}\right)k^{2m}
\end{multline}
\begin{equation}
 k\sin(l_p u k)= \sum_{s=0}^{\infty}\frac{(-1)^s(l_p u)^{2s+1}}{(2s+1)!}\,k^{2s+2}, 
\end{equation}
we obtain
\begin{multline}
    \Delta(u)\!=\!-\frac{1}{\varkappa^2+k_3^2}\! \mathop{\mathrm{res}}\limits_{k=0} \!\Bigg(\!\sum_{n=0}^{l-1}\!\sum_{m,s=0}^{\infty}\!\frac{(2n)!(n+l)!}{4^n(n!)^2(l-n-1)!}\! \times\!\\
    \times \Big(\frac{1}{k_3^{2m+2}}+\frac{(-1)^m}{\varkappa^{2m+2}}\Big)\frac{(-1)^s(l_p u)^{2s+1}}{r_0^{2n+3}(2s+1)!}\frac{1}{k^{2(n-m-s)+1}}\Bigg).
\end{multline}
Selecting terms which are simple poles in $k$ and calculating the residue, we obtain $\Delta(u)$ given by Eq.(\ref{residue-Delta}) in the main text.

\bibliographystyle{JHEP}
\bibliography{bibliography}

\end{document}